\documentclass[allclo]{FBSart}
\usepackage{amsfonts}
\usepackage{amssymb}
\usepackage{amsfonts}
\usepackage{amssymb}
\usepackage{amsthm}
\usepackage[normalem]{ulem}
\usepackage{epsf}
\usepackage[usenames]{color}
\newcommand{\cd}{\makebox[0.08cm]{$\cdot$}}

     \font\tenbifull=cmmib10 scaled 1200 
     \font\tenbimed=cmmib9
     \font\tenbismall=cmmib7
\mathchardef\bbkappa="7114 \mathchardef\bbgamma="710D
\mathchardef\bbrho="711A \mathchardef\bbsigma="711B
\mathchardef\bbtau="711C \mathchardef\bbvarrho="7125
\mathchardef\bbvarsigma="7126 \mathchardef\bbxi="7118

\usepackage[dvips]{graphicx}

\runningtitle{Three different approaches to the same interaction: the Yukawa model in nuclear physics}
\runningauthor{J.~Carbonell, F. de Soto and V.A.~Karmanov} 
\begin{document}

\newcommand{\beq}{\begin{eqnarray}}
\newcommand{\eeq}{\end{eqnarray}}
\newcommand{\be}{\begin{equation}}
\newcommand{\ee}{\end{equation}}
\newcommand \VEV [1] {\left\langle{#1}\right\rangle}

\title{Three different approaches to the same interaction: the Yukawa model in nuclear physics\thanks{Dedicated to Professor Henryk Witala at the occasion of his 60th birthday, June  2012}}
\author{J.\ Carbonell$^{a}$, F. de Soto$^{b}$ and V.A.\ Karmanov$^{c}$}
\institute{$^a$Institut de Physique Nucl\'eaire, Universit\'e Paris-Sud, IN2P3-CNRS, 91406 Orsay Cedex, France \\
$^b$Dpto. Sistemas F\'{\i}sicos, Qu\'{\i}micos y Naturales; U. Pablo de Olavide, 41013 Sevilla, Spain\\
$^c$Lebedev Physical Institute, Leninsky Prospekt 53, 119991 Moscow, Russia}

\maketitle
\begin{abstract}
After a brief discussion of the meaning of the  potential in quantum mechanics,
we shall examine the results of the Yukawa model (scalar meson exchange) for the nucleon-nucleon interaction in three different dynamical frameworks:
the non-relativistic dynamics of the Schrodinger equation,
the relativistic quantum mechanics of the Bethe-Salpeter and Light-Front equations
and the lattice solution of the Quantum Field Theory, obtained in the quenched approximation.
\end{abstract}

\section{Introduction}\label{intr}

Since Newton's time, an interaction between two particles is understood as something that prevents their relative motion to be rectilinear and uniform.
In non-relativistic Quantum Mechanics this is realized by any operator $\hat{V}$,  traditionally called potential,  that disturbs a plane wave $|k\rangle$, i.e.
there is no  any constant $\lambda$  such that
\[ (H_0+\hat{V}) \mid k\rangle = {\lambda} \mid k\rangle \]
while
\[ H_0 \mid k\rangle = E \mid k\rangle  \]
$E$ being the energy of the state $|k\rangle$.

From where does the potential $\hat{V}$  come from?
Sometimes it is taken from classical mechanics, like in the Coulomb case, but it can
be   {picked out of a hat} as well, provided one respects some space-time (translation, rotation, P, T) or internal (C, isospin) symmetries.

This way to built an interaction  is perfectly legal and can be even extended in order to satisfy the requirements of  relativistic invariance.
The  proper   method of constructing the Poincare algebra  with a given $\hat{V}$  was formulated in the series of works by Bakamjiam and Thomas \cite{BT_53}, extended later by
Keister and Polyzou \cite{KP_91}.

From this point of view, particles interact because they go into an inhomogeneous region of the space: if $\hat{V}$ {was} a constant, nothing would happen. This ``ex nihilo'' approach, though leading to some remarkable success, has not been very fertile when describing the physical phenomena, specially in the subatomic world.
Here is the crucible where deep changes in the states of matter occur,
one of the deepest  being the non-conservation of the particle number,
like in the annihilation of matter into light, $e^++e^-\to \gamma+\gamma$,
or, conversely,  the transform of ``speed'' into matter-antimatter pairs like in $p+p\to p+p+ \bar{p} +p $, both processes  which are nowadays customary in the laboratories. The potential approach of the non-relativistic quantum mechanics with a fixed number of particles is here of a little help.
Even worst, it does not provide any ``way of thinking'' about the natural processes.

A more interesting approach is provided by the Quantum Field Theory (QFT), now customary in the high energy physics world. In  QFT, the interaction is  a consequence of the exchange
of a bosonic mediator field. {This approach has been successfully applied since the development of quantum electrodynamics to every piece of the Standard Model of particle physics.} In the Lagrangian formulation, the simplest case is provided by the Yukawa model \cite{Yukawa_35}.
The interaction  between fermionic matter fields $\Psi$  is mediated by a scalar  field $\Phi$ and   written in terms of the Lagrangian as:
\[ {\cal L}(x)={\cal L}_D[\Psi] + {\cal L}_{KG}[\Phi] + {\cal L}_{int}[\Psi,\Phi]\]
where:
\[ {\cal L}_{int}(x) = g \bar\Psi(x) \;\Phi(x)\;\Psi(x)\]
and  ${\cal L}_D$,   ${\cal L}_{KG}$  are respectivley the Dirac and Klein-Gordon free Lagrangians.

\bigskip
Within this framework, a particle on a free state $|k_1\rangle$ emits a quanta $|q\rangle$ which is absorbed by a particle on a state $|k_2\rangle$.
Their initial states are modified in the process  and results into new ones $|k'_1\rangle$  and $|k'_2\rangle$ :  they have ``interacted''.
\begin{figure}[h!]
\begin{minipage}[h!]{8.5cm}
\begin{center}
\includegraphics[width=6.cm]{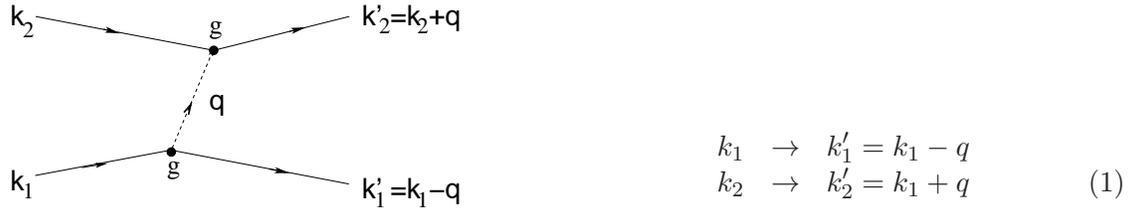}
\caption{Lowest order exchange graph.}\label{Exchange}
\end{center}
\end{minipage}
\begin{minipage}[h!]{7.5cm}
\begin{eqnarray}
k_1  &\to&k'_1 = k_1 - q \cr
k_2  &\to&k'_2= k_1 +q
\end{eqnarray}
\end{minipage}
\end{figure}

The basic element of this ``lego'' is the annihilation-creation-creation in the interaction vertex, driven by a strength constant $g_0$:
\[  g \; a^{\dagger}_{k'_1}\; b^{\dagger}_{q}\;a_{k^{\phantom{'}}_1} \quad\rightarrow\quad
g \; a^{\dagger}_{k'_1}e^{-ik'_1x_1}\;b^{\dagger}_{q}e^{iqx_1}\;a_{k_1}e^{ik_1x_1} \quad\rightarrow\quad
g \; \bar \Psi(x_1)\Phi(x_1)\Psi(x_1)\]
and Quantum Field Theory tells us how to associate to the process displayed in Fig. \ref{Exchange} a probability amplitude \( A(k_1,k_2\to k'_1,k'_2) \).

The relation between the QFT and the potential approach is made by identifying
$V$ to the amplitude of the lowest order ``exchange'' graph, which according to Feynman  rules reads
\begin{equation}
V\equiv \frac{1}{4m^2}A(k_1,k_2\to k'_1,k'_2) =  \frac{1}{4m^2}\bar{u} (k_1) u(k'_1) \; {g^2\over q^2-\mu^2} \; \bar{u} (k_2) u(k'_2)  \ .
\end{equation}

The {potential} $V(\vec{r})$  used in non-relativistic quantum mechanics is obtained
by a Fourier transform of the amplitude $A$ after some -- quite crude -- simplifications:
\begin{eqnarray*}
\bar{u} (k_1) u(k'_1) &=& \bar{u} (k_2) u(k'_2)=2m  \cr
q&=&(0,\vec{q}) =(0,\vec{k}_1-\vec{k}_2)  \cr
\end{eqnarray*}
One thus obtains
\[  V(k_1,k_2,k'_1,k'_2)=V(q)= {-} {g^2\over \mu^2 +   \vec{q}^2} \]
and, by a three-dimensional Fourier transform, the usual Yukawa  potential
\begin{equation} \label{VYuk_r}
V(r)= - {g^2\over 4\pi} \; {e^{-\mu r}\over r}
\end{equation}
is obtained. This is the procedure leading to the usual $1/r$ Coulomb potential
starting from QED. The same that brought Yukawa to  formulate the first theory of strong interactions  that deserved him a Noble prize.

\bigskip
Some remarks are in order:

{\it (i)} If the interaction was limited to only one exchange of Fig. \ref{Exchange}, there would never exist a bound state.
These states are indeed associated to poles in the scattering amplitude, and the Born term (one exchange term) has no one.
An infinite sum of exchanges is needed to generate these singularities and this task is ensured by the dynamical equations
\begin{figure}[h!]
\begin{center}
\includegraphics[width=6.cm] {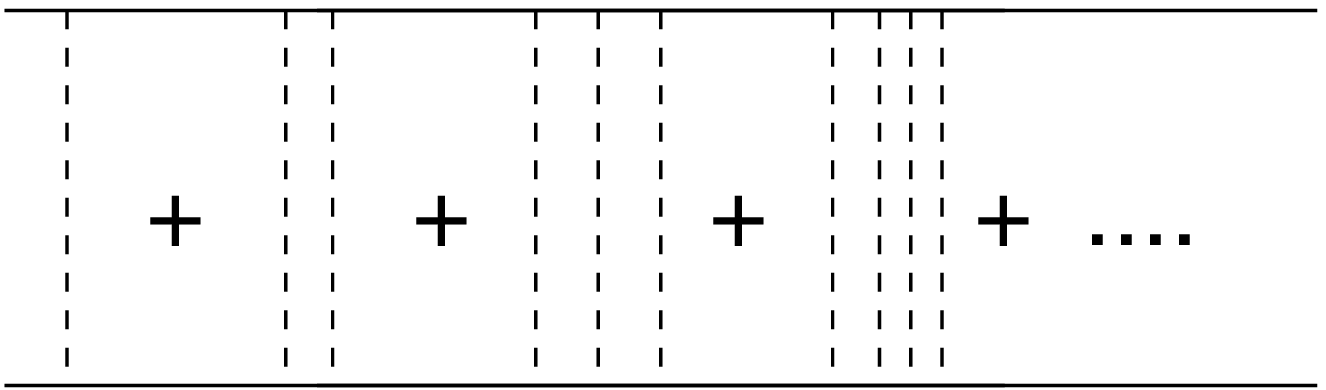}
\vspace{-0.5cm}
\end{center}
\end{figure}
\begin{equation}\label{LS}
 (H_0+V) |\Psi\rangle = E    |\Psi\rangle   \quad\Longrightarrow\quad   T= V+ VG_0 T = V + VG_0V+ VG_0VG_0V+ \ldots
 \end{equation}
Series (\ref{LS}) generates  a pole in the scattering amplitude  according
to the well known recipe: $ 1+x +x^2 + \ldots = {1\over1-x}$, named ``ladder sum''.

\begin{figure}[h!]
\begin{center}
\includegraphics[width=7.cm]{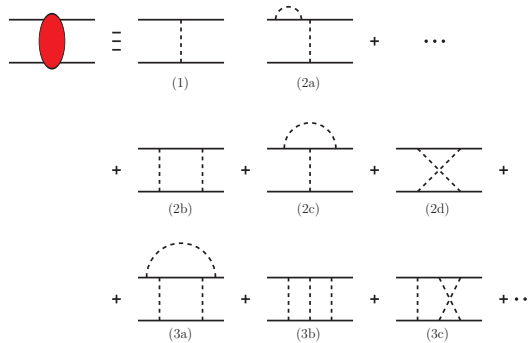}
\caption{Perturbative expansion of the scattering amplitude.}\label{all_diagrams.eps}
\end{center}
\end{figure}

{\it (ii)}  The ladder sum, used in all nuclear models,  accounts  only for a small, though infinite, part of the full interaction given by ${\cal L}_{int}$.
At any order in the coupling constant,  there are  several contributions which are ignored.
Figure  \ref{all_diagrams.eps}  illustrates the perturbative series of the scattering amplitude:
diagrams (2a), (2c) and (2d) are of the same order than (2b) but are not included in Eq. (\ref{LS}). It is even worse at higher orders, since the fraction of dropped diagrams increases very fast.

Even assuming that the effect of some of  the neglected diagrams
{is} incorporated in the renomalized quantities -- e.g. (2a) in the mass and (2c) in the coupling constant -- there
remains an infinity of them.  If $g$ is large, what one neglects is not negligible and one can expect that the physics with $\hat{V}$
could seriously depart from the ones contained in the underlying ${\cal L}_{\rm int}$.

\bigskip
We  present in what follows the results  obtained when the very same interaction --  the Yukawa model -- is considered
in three different dynamical approaches. They will be limited to the $J^{\pi}=0^+$ bound states and corresponding low energy scattering parameters.

Apart from being at the origin of the theory of  nuclear forces, this model,  the simplest  meson-fermion interaction Lagrangian, has several advantages:
{\it (i)} in the non-relativistic limit it gives the same result for the two-fermion and the two-boson system,
{\it (ii)} when inserted in the relativistic equations -- at least the ones considered here -- it does not require any regularization procedure to be integrated and
{\it (iii)}  it is a renormalizable quantum field theory.

Section 2 is devoted to the non-relativistic results. They are widely used  and constitute the reference ground   of  most of the nuclear
and atomic physics calculations.
In section 3 we will consider the results of two relevant, relativistic equations: the Bethe-Salpeter \cite{Salpeter:1951sz,GellMann:1951rw} and Light-Front Dynamics \cite{Carbonell:1998rj}.
Section 4 will contain the quantum field  results obtained using the lattice techniques in the quenched approximation.

\section{Non-relativistic results}

We consider in this section the non-relativistic system of two particles with equal mass $m$,
interacting  by a Yukawa potential (\ref{VYuk_r}) of strength $g$ and range parameter $\mu$.

Although the problem depends on three parameters ($m,g,\mu$) it can be shown that
the binding energy ($B$) and the scattering length ($a_0$) are given by
\begin{eqnarray}
 B &=& m \;\left(\frac{\mu}{m}\right)^2 \varepsilon(G) \label{B} \\
 a_0 &=& {1\over\mu} \; \lambda(G)
 \end{eqnarray}
where  $\varepsilon(G)$ and $\lambda(G)$ are respectively the
binding energy and scattering length of the dimensionless   S-wave  Schrodinger equation:
\begin{equation}\label{Sch}
u''(x) + \left[ -\varepsilon + G \;{e^{-x}\over x} \right] u(x)=0
\end{equation}
with a coupling constant $G$, related to the original parameters ($m,g,\mu$) by
\[  G = {g^2\over4\pi}\; {m\over\mu} \]

\begin{figure}[h!]
\begin{center}
\begin{minipage}[h!]{7.5cm}
\includegraphics[width=7.cm]{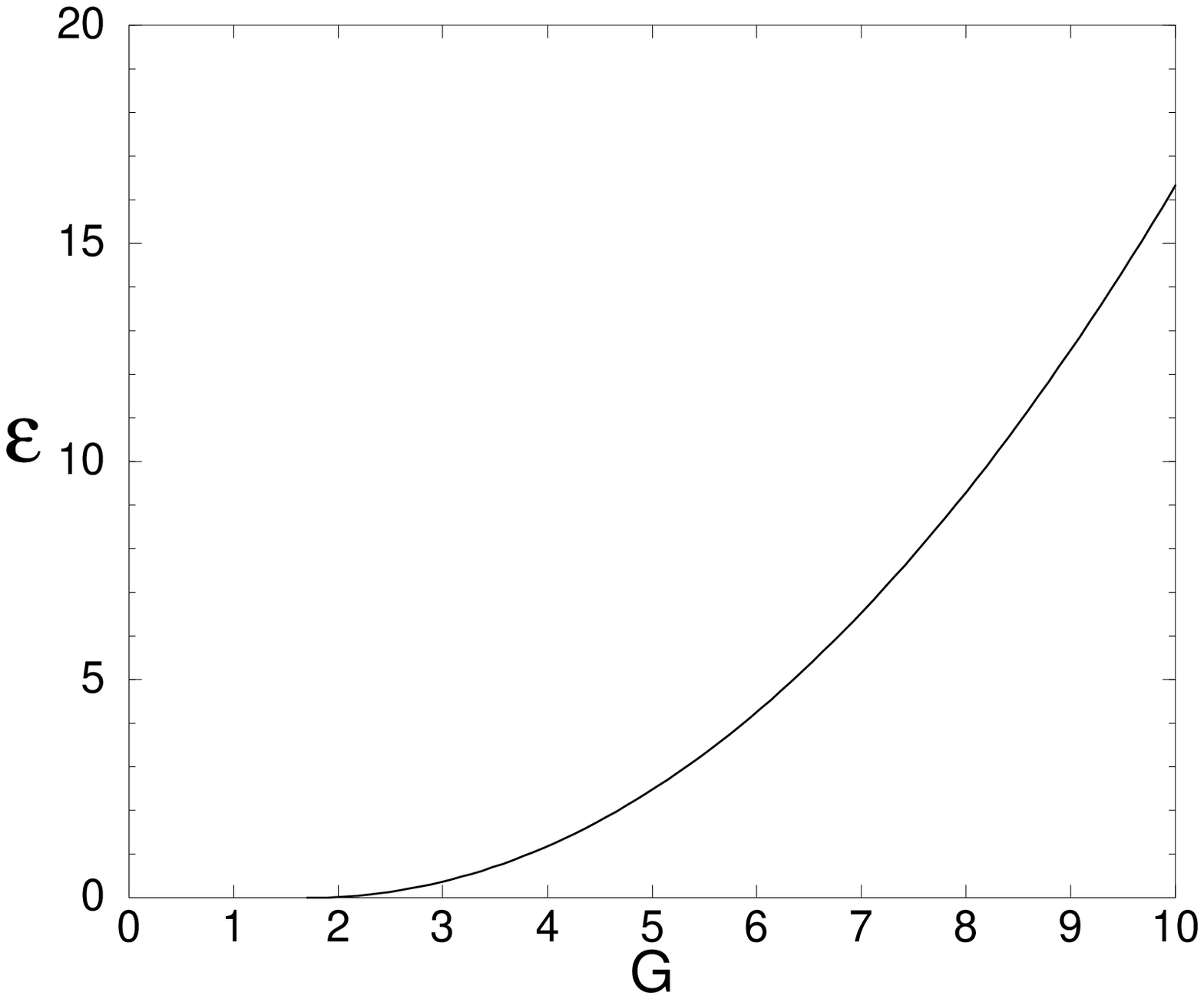}
\caption{Ground state binding energy of the dimensionless non-relativistic  Yukawa model (\ref{Sch}) as a function of the coupling constant $G$.
The appearance of the first bound state corresponds to $G_0=1.680$.}\label{eps_G}
\end{minipage}
\hspace{0.5cm}
\begin{minipage}[h!]{7.5cm}
\includegraphics[width=6.5cm]{Lambda_G_0_5.eps}
\caption{Scattering length   in the same model than Fig. \ref{eps_G}.
The Born approximation is indicated by the dashed line.
The singularity (vertical line) corresponds to the appearance of the first bound state.}\label{lambda_G}
\end{minipage}
\end{center}
\end{figure}

The functions $\varepsilon(G)$ and $\lambda(G)$ are displayed  in Figs. \ref{eps_G} and  \ref{lambda_G} respectively.
The convention used for the scattering length corresponds to  $\delta(k)=-a_0k +o(k^2)$.
The critical value for the appearance of the ground state is $G_0\approx1.680$.
At this value $\lambda(G)$ has a pole and it can be shown that for small  values of G one has
\begin{equation}\label{Born}
\lambda(G) = -G +o(G^2)
\end{equation}
which corresponds to the Born approximation.
All the physics of the non-relativistic problem (S-wave) is contained in these two figures,   with the understanding that
there exists an infinity of  similar branches  in $B(G)$ --   and corresponding poles in $\lambda(G)$ -- at increasing values of G  ($G_0=1.680,G_1=6.445,G_2=14.34,\ldots$) indicating
the appearance of an infinite number of excited states.

In summary,  for a massive exchange $\mu$,
there is a non-zero minimal value of the coupling constant $g_0$ required to have the first bound state.
It is given by ${g_0}^2(m,\mu)=4\pi G_0 {\mu\over m}$ with $G_0\approx 1.680$. This  $g_0$ value   decreases linearly with $\mu$
and vanishes in the Coulomb limit ($\mu\to0$) but in the nuclear case ($\mu/m\approx 0.5$) is rather large $g_0\approx 3$.
Once the bound state appears, the solutions of Eq. (\ref{Sch})  exist for any value of the parameter $G$ and one can obtain any value for the binding energy $B$, even exceeding $2m$,

\section{The Yukawa model in Relativistic Dynamics}

Things become less simple when considering the same model in a relativistic framework.
This  covers a very wide and --to some extent-- not well defined domain of theoretical physics
aiming to incorporate all or  part of the relativistic invariance in the dynamical equations.
It goes from the simple implementation of relativistic kinematics to the full Quantum Field Theory treatment, which will be the proper
way to incorporate relativity to the quantum world
but whose solutions are very difficult to obtain beyond the perturbative domain.

We will consider here two of the many relativistic approaches: the Light Front Dynamics (LFD) and the Bethe-Salpeter (BS)  equation.
Although far from being representative of this vast domain they illustrate well  the kind of
qualitative agreements and quantitative differences in the predictions they give.
Both are rooted in the Quantum Field Theory but present important differences
in the way they are constructed as well as in the formal objects they deal with.

\bigskip
The BS equation deals with a field theoretical object  corresponding to the following amplitude \cite{GellMann:1951rw}: 
\begin{equation}\label{BSA}
\Phi(x_1,x_2,P)=\langle 0|T\left\{\Psi(x_1)\bar\Psi(x_2)\right\}|P\rangle
\end{equation}
where $P^2=M^2$ is the total mass of the two-body system.

It is usually written in momentum space $\Phi(k,P)$, obtained after taking into account translational invariance and performing a Fourier transform with respect to
 the relative coordinate  $x=x_1-x_2$:
\[ \Phi(x_1,x_2,P)=\frac{1}{(2\pi)^{3/2}} \; \tilde{\Phi}(x,P) \;e^{-iP \cd (x_1+x_2)/2}  \]
$$ \tilde\Phi(x,P)= \int \frac{d^4x}{(2\pi)^4} \;\Phi(k,P)\; e^{-ik\cd x}. $$

The BS amplitude for a two-fermions system is  a $4\times 4$ matrix in spinor space
which can be expanded in a basis of independent Dirac structures $S^{(c)}$
\begin{equation}\label{Phi_BS}
 \Phi =\sum_{c=1}^{n_c} \phi_c \; S^{(c)} \qquad   S^{(c)}\in \left\{ 1, \gamma_{\mu}, \gamma_5 , \gamma_{\mu}\gamma_5 , \sigma_{\mu\nu} \right\}
 \end{equation}
The number of scalar components $n_c$ depends on the quantum number of the state.
For $J=0^+$ there are four of them which can be chosen as:
\begin{equation}
\Phi(k,P) =
  \phi_1(k,P)\gamma_5
+ \phi_2(k,P){\hat{P}\over M}\gamma_5
+ \phi_3(k,P)\left[{(k\cd P)\hat{P}\over M^3}-{\hat{k}\over M}\right]\gamma_5
+ \phi_4(k,P){i\sigma_{\mu\nu}P_{\mu}k_{\nu}\over M^2}
\end{equation}
where $\hat{k}=\gamma^{\mu}k_{\mu}$ and $ \sigma_{\mu\nu} ={i\over2} [\gamma_{\mu},\gamma_{\nu}] $.

The dynamical equation can be represented   in the following graphical form:
\setlength{\unitlength}{1.2mm}
\begin{picture}(80,40)(5,0)  
\put(40,21){\line(-3,+1){20}} \put(25,15.0){\vector(3,+1){2}} \put(15,28){$k_1$}
\put(40,20){\line(-3,-1){20}} \put(25,26.0){\vector(3,-1){2}} \put(15,12){$k_2$}
\put(40,20){\line(10,0){15}}
\put(40,21){\line(10,0){15}}
\put(40,20.5){\circle*{2}} \put(47,16){$p$}

\put(62,19.5){=}\put(70,19.5){\tiny{$iK(k_1,k_2,k'_1,k'_2)$}}

\put(75,27.7){\line(10,0){15}}
   \put(80,27.7){\vector(1,0){2}}\put(70,28){$k_1$}\put(90,27.5){\circle*{1}}\put(89,29){$\Gamma_1$}\put(100,26){$k'_1$}
\put(75,13.4){\line(10,0){15}}
   \put(80,13.3){\vector(1,0){2}}\put(70,12){$k_2$}\put(90,13.2){\circle*{1}}\put(89,9){$\Gamma_2$}\put(100,13){$k'_2$}
\put(90,13.4) {\dashbox{0.5}(0,14.3)[t]}   
\put(110,21){\line(-3,+1){20}} \put(95,15.0){\vector(3,+1){2}}
\put(110,20){\line(-3,-1){20}} \put(95,26.0){\vector(3,-1){2}}
\put(110,20){\line(10,0){15}}
\put(110,21){\line(10,0){15}}
\put(110,20.5){\circle*{2}}
\end{picture}

For the case of two equal mass fermions it reads
\begin{equation}\label{EBSM_F}
S_1^{-1}(k_1)\Phi(k,P)\bar{S}_2^{-1}(k_2) = \int \frac{d^4k'}{(2\pi)^4}\;  iK(k,k')\;  \Gamma_1\;\Phi(k',P)\;\bar{\Gamma}_2
\end{equation}
where
\begin{eqnarray*}
S_1(k_1)      &=&\frac{i}{\hat{k}_1-m+i\epsilon}  \;\;\; ,\qquad\qquad   k_1={P\over2}+ k \cr
\bar{S}_2(k_2)&=&\frac{i}{-\hat{k}_2-m+i\epsilon} \;,\qquad\qquad   k_2={P\over2}- k
\end{eqnarray*}
are the fermion propagators. $\Gamma_i$ and $K$ are respectively the vertex functions and the meson propagator.
They depend on the particular type of meson-fermion coupling.
In the case of the Yukawa model (scalar meson exchange)  they are
\begin{equation} \label{ladder}
K(k,k',p)= \frac{1}{(k-k')^2-\mu^2+i\epsilon}.
\end{equation}
and
\[\Gamma_1=\Gamma_2=\bar{\Gamma}_2=-ig\]

The BS equation (\ref{EBSM_F}) in momentum space is  four dimensional. After a partial wave expansion
it reduces to a set of coupled two-dimensional equations among the different components $\phi_c$ of the state (\ref{Phi_BS}).
This equation has been solved by several authors for the bound state problem both in the Euclidean  \cite{Dorkin:2007wb,Dorkin:2010dj}
and Minkoswki metric \cite{Carbonell:2010tz,Carbonell:2010zw}
and for
the on-mass shell scattering amplitudes \cite{FT_NPB84_75,Fleischer:1977yd,Fleischer:1980qr}.
For the scattering states, the full (off-shell) solution in Minkowski space has been obtained only very recently \cite{KC_LCM_2012,KC_FB20_2012,KC_BALDIN_2012}.

\newcommand{\dpp}{{\partial _p}} \newcommand{\dL}{\partial{\cal L}} \newcommand{\dmuphi}{\partial_{\mu}\phi}
\newcommand{\dNuphi}{\partial^{\nu}\phi} \newcommand{\dnuphi}{\partial_{\nu}\phi}
\newcommand{\dLdmuphi}{\partial{\cal L}\over\partial(\partial_{\mu}\phi)}

\bigskip
LFD  can be understood as an Hamiltonian
formulation of the QFT defined on an space-time surface of equation  $\omega \cd x=\sigma$, where
$\omega$ is a light cone vector $\omega^2=0$  \cite{Carbonell:1998rj}.

The state vector $|\Psi(\sigma)\rangle$ is defined on this plane and the Poincar\'e algebra generators are obtained
by integrating through this surface the flux of the conserved  Noether currents associated to a given Lagrangian ${\cal L}$.
In case of translations, for instance, they are given by
\begin{equation}\label{GLGD}
 \hat{P}^{\mu}(\sigma)=\int T^{\mu\nu}(x)\delta(\omega\cd x-\sigma) \;\omega_{\nu}d^4x
\end{equation}
with
\[T^{\mu\nu}={\dLdmuphi}{\dNuphi}-g^{\mu\nu}{\cal L}\]
and fulfill
\[ \partial_{\sigma}\hat{P}^{\mu}(\sigma) =0 \]

Once the generators are obtained, the dynamical equation determining the mass of the system $M^2$ is given by
\begin{eqnarray}
\hat{P}^2 \mid\Psi\rangle &=&  M^2 \mid\Psi\rangle \label{P2}
\end{eqnarray}
After some algebra, one is led to
\begin{equation}\label{eqlfd0}
\fbox{$\displaystyle (M^2 - P_{0}^{2}) \mid\Psi\rangle = 2 P_{0}\cd\omega \; \int \;
{\cal{H}}_{int}(\omega \tau) \exp(-i \sigma \tau) \; d \tau \mid\Psi\rangle  $}
\end{equation}
where ${\cal{H}}_{int}(k)$ denotes the Fourier transform of the hamiltonian density
\begin{eqnarray*}
{\cal{H}}_{int}(k)=\int {\cal{H}}_{int}(x) \exp(ik\cd x) d^4x
\end{eqnarray*}

The state vector is decomposed into its Fock components  with an increasing number of particles,
which can be schematically written as:
\begin{eqnarray}
\mid\Psi\rangle &=& \sum_{\alpha\beta}\int d^4k_1\ldots d^4k_{\alpha}d^4q_1\ldots d^4q_{\beta} \cr
&&
\Psi_{\alpha\beta}(k_1,\ldots,k_{\alpha},q_1\ldots q_{\beta})  \;
   a^{\dagger}_{k_1} \ldots a^{\dagger}_{k_\alpha} \;
   b^{\dagger}_{q_1} \ldots b^{\dagger}_{q_\beta}  \;\mid0\rangle    \label{WF}
\end{eqnarray}
The components of this expansion $\Psi_{\alpha\beta}$ are the relativistic counterparts of the usual non-relativistic wave functions.
They have also a probability interpretation and are smooth functions of the arguments.
We can consider the set
$\Psi_{n_{\alpha\beta}}\equiv\{\Psi_{\alpha\beta}\}$ as the
components of an infinite dimensional vector $\Psi= (\Psi_1,
\Psi_2, \Psi_3, \dots)$, coupled to each other via the interaction operator ${\cal{H}}_{int}$.
Equation (\ref{eqlfd0}) is thus an infinite system of coupled channels.
\begin{eqnarray*}
(M^2-P_0^2) \pmatrix{\ldots\cr \Psi_2 \cr \Psi_3\cr \ldots } = 2 P_{0}\cd\omega \;  \int \; {\cal{H}}_{int}(\omega \tau) d \tau \pmatrix{ \ldots \cr\Psi_2 \cr \Psi_3 \cr \ldots}
\end{eqnarray*}
If we restrict ourselves to the two-
($\Psi_{2}\equiv\{\Psi_{20}\}$)  and three-body
($\Psi_{3}\equiv\{\Psi_{21}\}$) wave functions, we obtain a system
of two coupled equations for $\Psi_2$ and $\Psi_3$ which
constitutes the ladder approximation. By expressing $\Psi_3$ in
terms of $\Psi_2$, one gets an integral equation for $\Psi_2$ with an energy-dependent kernel.

The LF equation for the  two-fermion system is three-dimensional and takes the form
\begin{equation}\label{LFE}
\left[M^2-4(\vec{k}\,^2+m^2)\right]{\mit\Phi}(\vec{k},\vec{n})=
\frac{m^2}{2\pi^3} \int  K(\vec{k},\vec{k}\,',\vec{n},M^2)
{\mit \Phi}(\vec{k}\,',\vec{n}) \frac{d^3k'}{\varepsilon_{k'}}
\end{equation}
where $K(\vec{k},\vec{k}\,',\vec{n},M^2)$ is the interaction kernel and $\vec{n}$ is unit vector, the spacial part of  the light cone one $\omega=(\omega_0,\omega_0\vec{n})$.
For the Yukawa model it reads
\begin{equation}\label{eq6}
K ( \vec{k},\vec{k}\,',\vec{n},M^2) =-\frac{g^2}{4m^2(Q^2+\mu^2)} \left[\bar{u}(k_2) u_{\sigma'_2}(k'_2)\right]\,  \left[\bar{u}(k_1) u_{\sigma'_1}(k'_1)\right],
\end{equation}
with
\begin{eqnarray}\label{Q2}
 Q^2 &=& (\vec{k}-\vec{k'})^2
-(\vec{n}\cd\vec{k})(\vec{n}\cd\vec{k'})
{(\varepsilon_{k'}-\varepsilon_{k})^2\over\varepsilon_{k'}\varepsilon_k}
+\left(\varepsilon_{k'}^2+\varepsilon_k^2-{1\over2}M^2\right)\:\left|{\vec{n}
\cd\vec{k'}\over\varepsilon_{k'}}-{\vec{n}\cd\vec{k}\over\varepsilon_{k}}\right|
\end{eqnarray}

The $J=0^+$ wave function  has the form
\begin{equation}\label{Phi_LF_J0}
{\mit \Phi}(\vec{k},\vec{n})= \overline{u}(k_2) \phi U_c \overline{u}(k_1),
\end{equation}
where $\phi$ is expanded in terms of spin structures $S_i$
\begin{equation}\label{eq1_1}
\phi=f_1S_1 +  f_2 S_2,
\end{equation}
\begin{equation}\label{eq1_2}
S_1=\frac{1}{2\sqrt{2}\varepsilon_k}\gamma_5,\quad
S_2=\frac{\varepsilon_k}{2\sqrt{2}m  \mid \vec{k} \times \vec{n} \mid}
\left(\frac{2m\vec{\omega}}{\omega\cd p}-
\frac{m^2}{\varepsilon_k^2}\right)\gamma_5
\end{equation}
and $f_{i}$ are scalar components depending on $(k,\vec{k}\cd\vec{n})$ .

\bigskip
By inserting the expansion (\ref{Phi_LF_J0})  in (\ref{LFE}) it results -- like for the BS case -- in a system of  two-dimensional integral equations coupling
the different components of the wave function $\phi_i$
despite the fact that the LF equations is only three-dimensional. This {is} due to the existence of an additional vector $\vec{n}$ in the theory.
Notice however that the number of components in LFD is 2,  half the number in the BS case.
The explicit equations and the numerical solutions of the {LFD}  equation for the Yukawa model have been
presented in \cite{glazek1,glazek2,ManginBrinet:2001we,ManginBrinet:2001tc,Karmanov:2001te,ManginBrinet:2001md,Karmanov:2001mj,ManginBrinet:2003nm}.

\vspace{1.cm}
\begin{figure}[h!]
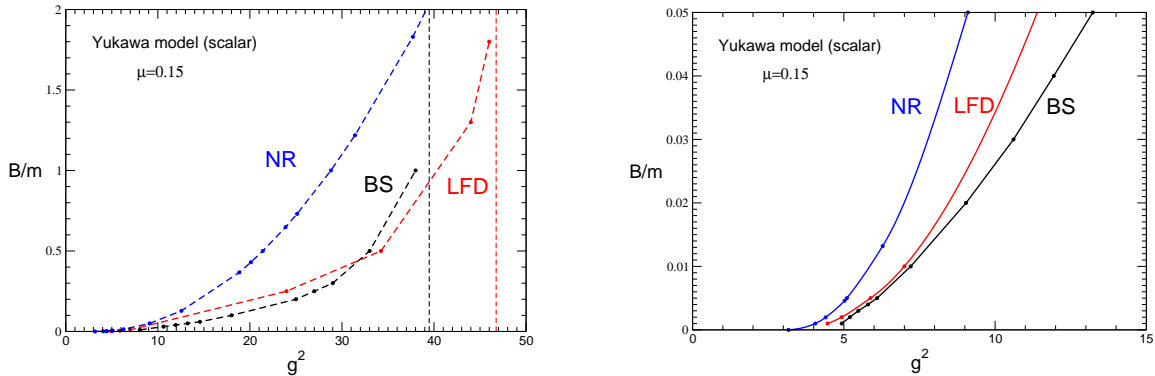

\begin{minipage}[h!]{7.5cm}
\begin{center}
\includegraphics[width=7.cm]{B_g2_S_SFF.eps}
\end{center}
\end{minipage}
\hspace{0.5cm}
\begin{minipage}[h!]{7.5cm}
\begin{center}
\includegraphics[width=7.cm]{Zoom_B_g2_S_SFF.eps}
\end{center}
\end{minipage}
\caption{Binding energy of the  $J=0^+$ state in Yukawa model as a function of the coupling constant $g^2$
given by the Light-Front (LF), Bethe-Salpeter (BS) and non-relativistic Schrodinger equation.
Vertical dotted line denotes the critical coupling constant $g_c=6.84$ for LF and $g_c=2\pi$ for BS.}\label{B_g2_S}
\end{figure}

The BS and LFD binding energy of the  $J=0^+$ state as a function of the coupling constant is displayed in Fig. \ref{B_g2_S}
for the parameters $m=1$ and $\mu=0.5$ and is compared to the non-relativistic results given by (\ref{B}).

It is worth noticing the existence, for both relativistic equations, of a critical coupling constant $g_c$.
Above that value, $g>g_c$, the systems ``collapses'', i.e. its spectrum is unbounded,
and vertex form factors are required to solve the corresponding dynamical equations.
Indeed, when $g\to g_c$ from below, the value of $M^2$ vanishes, becomes negative and tends smoothly to $-\infty$.
This happens in a rather narrow domain where $g$ is very close to $g_c$ and which is not very distinguished in Fig. \ref{B_g2_S}.
The physical meaning of  BS solution is lost already at  $g<g_c$ though very close to, when $M^2$ becomes negative.
This result was first found in \cite{ManginBrinet:2001we,ManginBrinet:2001tc,Karmanov:2001te} in the framework of  LFD
and also obtained for the BS equations \cite{Carbonell:2010zw}  using the methods
developed in \cite{ManginBrinet:2001tc}.

If the very existence of this critical coupling constant is common to both relativistic approaches, their precise numerical value, independent of $m$ and $\mu$,
however depends on the particular dynamics: one has $g_c=2\pi$ for BS and $g_c=6.84$ for LFD.
This difference is due to the different treatment of the intermediate states in the ladder kernel in these two approaches:
while the ladder BS equation incorporate effectively the so-called
stretch-box diagrams \cite{SBK_PRC58_1998} , they are absent in the ladder LFD results.

Contrarily to the non-relativistic case, the range of the strength parameter $g$ in these relativistic equations is  limited.
These limits are indicated by   vertical dashed lines in  Fig. \ref{B_g2_S}. As one can see the accessible binding energies
are the same in both equaztions $B\in[0,M]$ due to the vanishing of $M^2$ near $g_c$.

The results of both relativistic equations are quite close to each other, specially at moderate values of B (see right panel of Fig.  \ref{B_g2_S})
but  depart  from the non-relativistic ones which, for a given value of $g$, generate always much more attraction.
This differences are not of kinematical origin, since they exist even in the limit of zero binding energies  and increase with the value of $\mu$.
Strictly speaking the results would coincide only in the limits  $\mu\to0$ (Columb problem) and $g\to0$.

To our knowledge there are no  published results for the scattering observables with the  Yukawa model and the considered equations.
They have been however computed for the scalar  model ($\phi^2\chi$ theory) both in LF \cite{Ji:1992xr,Oropeza_PhD_2003} and
in the BS one \cite{Levine:1966zz,Levine:1967zza,KC_LCM_2012,KC_FB20_2012,KC_BALDIN_2012}.

\bigskip
The results described above illustrate well the kind of dispersion one can find when moving from a non-relativistic to a relativistic description of the same system,
would be the simplest one and submitted to the simplest interaction.
While the result of the non-relativistic Schrodinger  equation with a given potential is unique, it is not the case in the relativistic world.
The implementation of relativity  can be done following different approaches,
but they give rise to qualitative results which are common to some of them.
For instance: the strong repulsive effects, the existence of critical coupling constants
or, when solving the three body system,  the automatic generation of three-body forces \cite{Karmanov:2008cd,Karmanov:2008bx}.
The reason for such  differences  is not in kinematics.
Notice that, as a consequence of the inequality
\[   \sqrt{p^2+m^2} - m- {p^2\over 2m} <  0 ,  \]
the relativistic kinematical corrections  are always attractive, while the results of fig \ref{B_g2_S}, and similar one for the scalar theories,  shows rather a strong repulsion.
The origin of this new behavior  is thus dynamical  and lies in  the interaction kernel as it can
be seen by computing the zero energy cross sections (see Fig. 1 from  \cite{KC_FB20_2012}) .

\bigskip
One of the more consistent approaches to  relativistic {\it ab initio} nuclear physics is the one developped by Gross and Stadler
using the spectator equation \cite{SG}. The philosophy is quite close to the BS and LF equation:
using this relativistic equation and OBE kernels, these authors obtain
a very good fit to pn data with a relatively small number of parameters
and reproduce the experimental triton binding energy without explicitly adding 3-body forces.
We would however remark,  that there exists other relativistic
approaches which substantially differ from the ones described above.
Of particular interest is the approach developped by H. Kamada, W. Gloeckle, H. Witala, J. Golak, Ch. Elster, W. Polyzou and coworkers.
The starting point is a potential which in the non relativistic dynamics provides a satisfactory description of NN data.
Using  the Bakamjiam-Thomas construction  \cite{BT_53,Polyzou:2010kx}, a new relativistic potential is  obtained    in such a way that once inserted in a relativistic
Lipmann-Schwinger equation it produces the same phase shifts than the non relativistic ones \cite{Kamada:1999wy}.
The parameters of this new potential are not readjsuted: it is  an implicit function of the preceding ones and contains no new parameters.
In this scheme there are, by construction, no any two-body relativistic effects.
The real difference between relativistic and non relativistic dynamics appears only when going to the three-body problems.
This  approach has been succesfully applied to the few-nucleon problem  \cite{Witala:2005nw,Lin:2007ck,Kamada:2007ms,Elster:2008ca,Witala:2009zzb}.

\section{The Yukawa model in the Lattice}

The very large effects we found when including the cross ladder kernel in the BS and LF equations  \cite{Carbonell:2006zz},
as well as the pionneer work of  \cite{Nieuwenhuis:1996mc}  computing the full cross ladder sum in the scalar theories
motivated a work to evaluate the full QFT content of the Yukawa model.

A series of papers  \cite{deSoto:2005jj,deSoto:2006jr,deSoto:2006jt}   has been devoted to this project  with the aim
of obtaining the $B(g^2)$ dependence of the Yukawa model as well as some low energy parameters.
They are summarized in \cite{deSoto:2011sy}

To this aim we have used  the standard lattice techniques, developed in the framework of QCD, and  the Lagrangian  density:
\begin{equation}\label{lagrangian}
\mathcal{L}=\mathcal{L}_{D}(\bar\Psi,\Psi) + \mathcal{L}_{KG}(\Phi) + \mathcal{L}_{I}(\bar\Psi,\Psi,\Phi)\ ,
\end{equation}
with, in Euclidean space,
\begin{eqnarray}
\mathcal{L}_{D}(\bar\Psi,\Psi)      &=&  \overline\Psi\left(\partial_\mu\gamma^\mu + m_0\right) \Psi\ , \label{LD}\\
\mathcal{L}_{KG}(\Phi)              &=&  \frac{1}{2}\left(\partial_\mu\Phi\partial^\mu\Phi + \mu_0^2\Phi^2 \right) \ ,\label{LKG} \\
\mathcal{L}_{I}(\bar\Psi,\Psi,\Phi) &=&  g_0 \overline\Psi\Phi\Psi  
\end{eqnarray}
where $\Psi$ denotes respectively the fermion field and $\Phi$ the exchanged meson field responsible for the interaction.

The fermion field is supposed to describe a nucleon (N) and the meson field a -- more or less
fictitious -- scalar particle ($\sigma$) responsible for the attractive part of the NN potentials.
The Lagrangian depends on three parameters: the fermion $m_0$ and  meson $\mu_0$ masses and
a dimensionless coupling  constants $g_0$.

The theory is solved in a discretized space-time  Euclidean  lattice
of volume $V=L^3\times T$ and lattice spacing $a$ using the Feynman path integral formalism.
In this approach, the vacuum expectation values of an arbitrary operator $\mathbf{O}$  is given by the integral:
\begin{equation}\label{vev}
\VEV{\mathbf{O}(\bar\Psi,\Psi,\Phi)}\ =\ \frac{1}{Z} \; \int[d\bar\Psi][d\Psi][d\Phi] \mathbf{O}(\bar\Psi,\Psi,\Phi) \; e^{-S_E[\bar\Psi,\Psi,\Phi]}\ ,
\end{equation}
where, according to  (\ref{lagrangian}),  the discretized Euclidean action $S_E$ can be written  in the form
\[ S_E=a^4\sum_x \mathcal{L} = S_D + S_{KG} + S_I\]
and plays the role of a probability distribution in a Monte Carlo simulation.

The  fermionic part ($S_D$+$S_I$)  is  written as a bilinear form in the dimensionless fermion fields $\psi= \sqrt{a^3\over2\kappa} \Psi$:
\beq\label{S_D}
S_D +S_I = \sum_{xy} \bar\psi_x D_{xy} \psi_y
\eeq
where
\beq\label{DiracWilson}
D_{xy} = \delta_{x,y} - \kappa \sum_\mu \left[ \left(1-\gamma_\mu\right) \delta_{x,y-\mu} + \left(1+\gamma_\mu\right) \delta_{x,y+\mu} \right]   + g_L \phi
\eeq
is the Dirac-Yukawa operator,
\beq\label{kappa}
\kappa=\frac{1}{8+2am_0}
\eeq
 the hopping parameter and $g_L=2\kappa g_0$ the lattice coupling constant. In terms of the dimensionless meson field $\phi=a\Phi$, the discrete Klein-Gordon action reads:
\beq\label{S_KG}
S_{KG} = \frac{1}{2} \sum_{x} \left[ \left( 8 + a^2\mu_0^2 \right) \phi_x^2 - 2 \sum_{\mu} \phi_{x+\mu}\phi_x \right]
\eeq

The integral over the fermion fields in (\ref{vev}) is performed by algebraic methods.
The keystone in a lattice simulation is   the fermion propagator $S(x,y)$, corresponding to $\mathbf{O}(\bar\psi,\psi,\phi)=\psi_x\overline\psi_y$.
After performing the fermionic integration, one is left with
\begin{equation}\label{unquenched}
S(x,y)\ =\ \VEV{\psi_x\overline\psi_y}\ =\ \frac{1}{Z} \int[d\phi]\ D^{-1}_{xy}\ {\rm det} [D(\phi)]\ e^{-S_{M}(\phi)}\ ,
\end{equation}
This implies the evaluation of a determinant and inverse the Dirac operator that, even for
moderate lattices  $V\sim 24^4$, has a dimension of $\sim 10^6$. Moreover, if a Monte
Carlo simulation is to be done using  Eq. (\ref{unquenched}), the probability distribution
for meson configurations is given by  $e^{-S_M(\phi)-\log(\det(D))}$, what means
evaluating a large determinant in every Monte Carlo step . This can be avoided by the
use of Hybrid Monte Carlo techniques that nevertheless are the main source of time spent
in the simulation. This task is considerably  simplified in the ``quenched'' approximation
that, from  the computational point of view consists in setting $\rm{det(D)}$ independent
of  the meson field in the fermionic integral.

\begin{figure}[h!]
\vspace{-0.5cm}
\begin{center}
\includegraphics[width=6.cm]{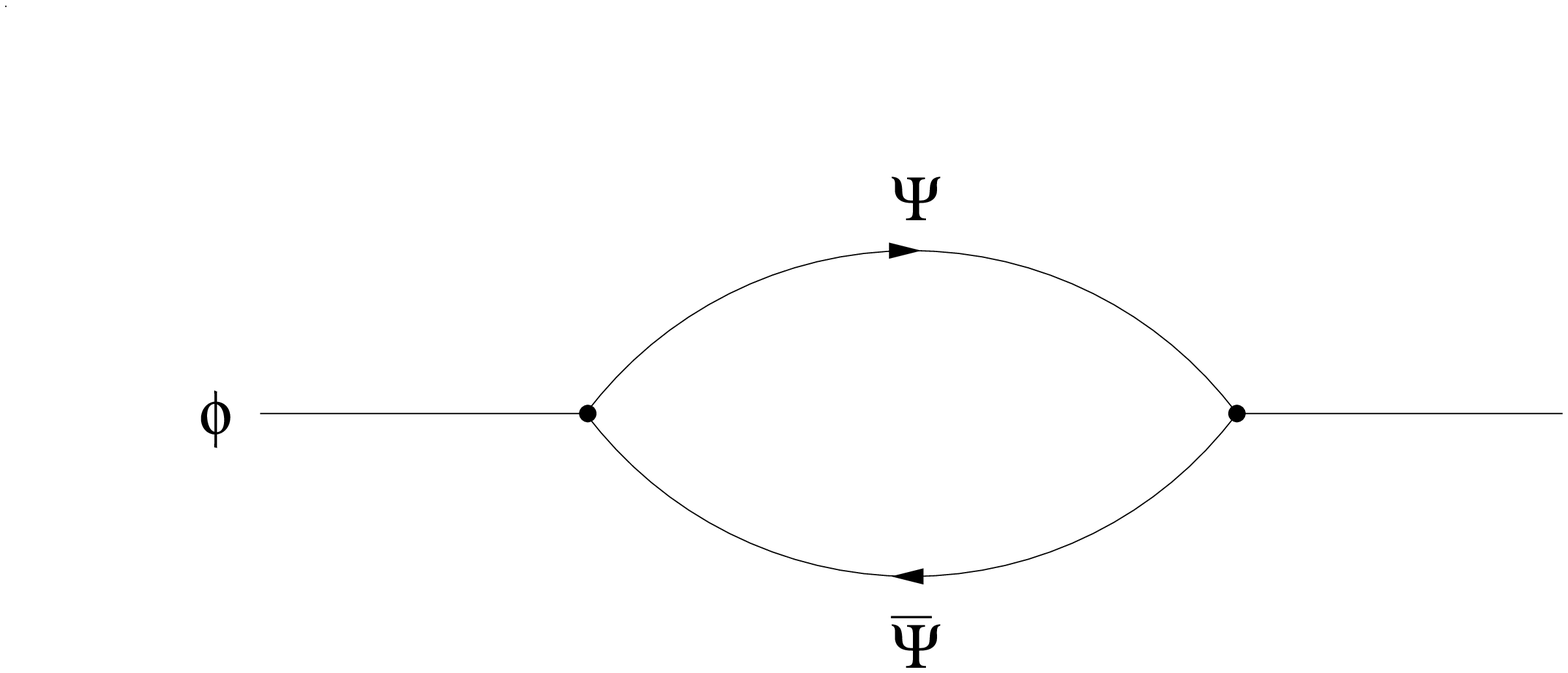}
\end{center}
\vspace{-0.5cm}
\vspace{.cm} \caption{The quenched approximation neglects the possibility
for a meson $\Phi$ to create a virtual fermion-antifermion pair $\Psi\bar\Psi$.} \label{quench}
\end{figure}

From a physical point of view, the quenched approximation avoids the possibility for a meson to create a
virtual  nucleon-antinucleon pair $\Phi\to\bar\Psi\Psi$ (see Fig. \ref{quench}).
Due to the heaviness of the nucleon with respect to the exchanged meson  this
approximation is fully justified in low energy nuclear  physics  and implicitly assumed  in all  the potential models.
Under this hypothesis the generation of meson-field configurations according to the probability distribution $e^{-S_M(\phi)}$ is also greatly simplified
and the path-integral sum over the mesonic fields in $\ref{unquenched}$ can be accurately computed.
to compute $D^{-1}_{x y}[\phi]$ for an statistical ensemble of meson field configurations.
\begin{equation}
S(x,y) \approx \ \frac{1}{N} \sum_{i=1}^N D^{-1}_{xy} (\phi_i)
\end{equation}
Due to translational invariance one is left in practice to compute $S(x,0)\equiv D^{-1}_{x 0}[\phi]$, that is to  solve the linear system:
\beq\label{DS}
D_{z x}(\phi) S_x = \delta_{z 0}
\eeq

One can obtain in this way the renomalized fermion mass $am$ as well as the mass of the two-body interacting particles $aM_2$, both in lattice units.
The binding energy -- in constituent mass units --  is then given by $B/m=(aM_2-2am)/am$.

In Fig. \ref{fig:B_vs_L} we show this binding energy
as a function of the lattice size $La\mu$ for a given set of parameters. The dotted line is a fit obtained with a $1/L^3$ dependence.
As it can be
seen in this figure, the binding tends to zero in the infinite volume limit. This indicates that this two-fermion system has no bound state
for this particular set of parameters.
The binding energy results from setting two interacting particle in a box with  periodic boundary conditions
but  contrary to a real bound state, this one disappears in the limit $L\to\infty$.
It turns out that the situation is however the same for the whole range of parameters accessible in the numerical simulations.

As in the relativistic dynamics, though for a completely different reason, there exist a maximum value of the coupling constant
that can be attained within this framework. The reason is the existence of  zero modes in the Dirac-Yukawa operator, i.e. the
appearance of meson field configurations such that  ${\rm det}[D(\phi)]\approx0$ thus leading to an ill-conditioned linear
system (\ref{DS}) and the impossibility to compute the fermion propagator.
As a practical measure of the ``ill-conditioness'' of $D$ we have considered its
``condition number'' defined as the ratio between the largest to the lowest eigenvalue
modulus. The largest is this number the more difficult is to solve the linear system.
Depending on the method used for that purpose, either the algorithm cannot find the
solution, or the round-off errors make the solution wrong.

\begin{figure}[h!]
\begin{center}
\includegraphics[width=8.cm]{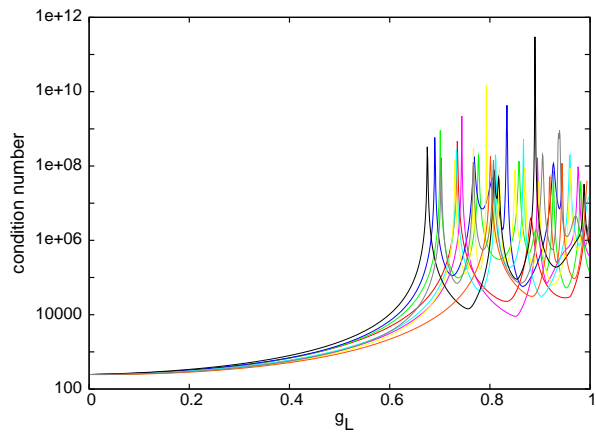}
\caption{Conditioning number as a function of $g_L$ for a fixed value of $\kappa=0.11$ and $V=8^4$ and for 9 different meson field configurations.} \label{fig:cn}
\end{center}
\end{figure}

\begin{figure}[h!]
\vspace{0.5cm}
\begin{center}
\begin{minipage}[h!]{7.cm}
\begin{center}
\includegraphics[width=6.9cm]{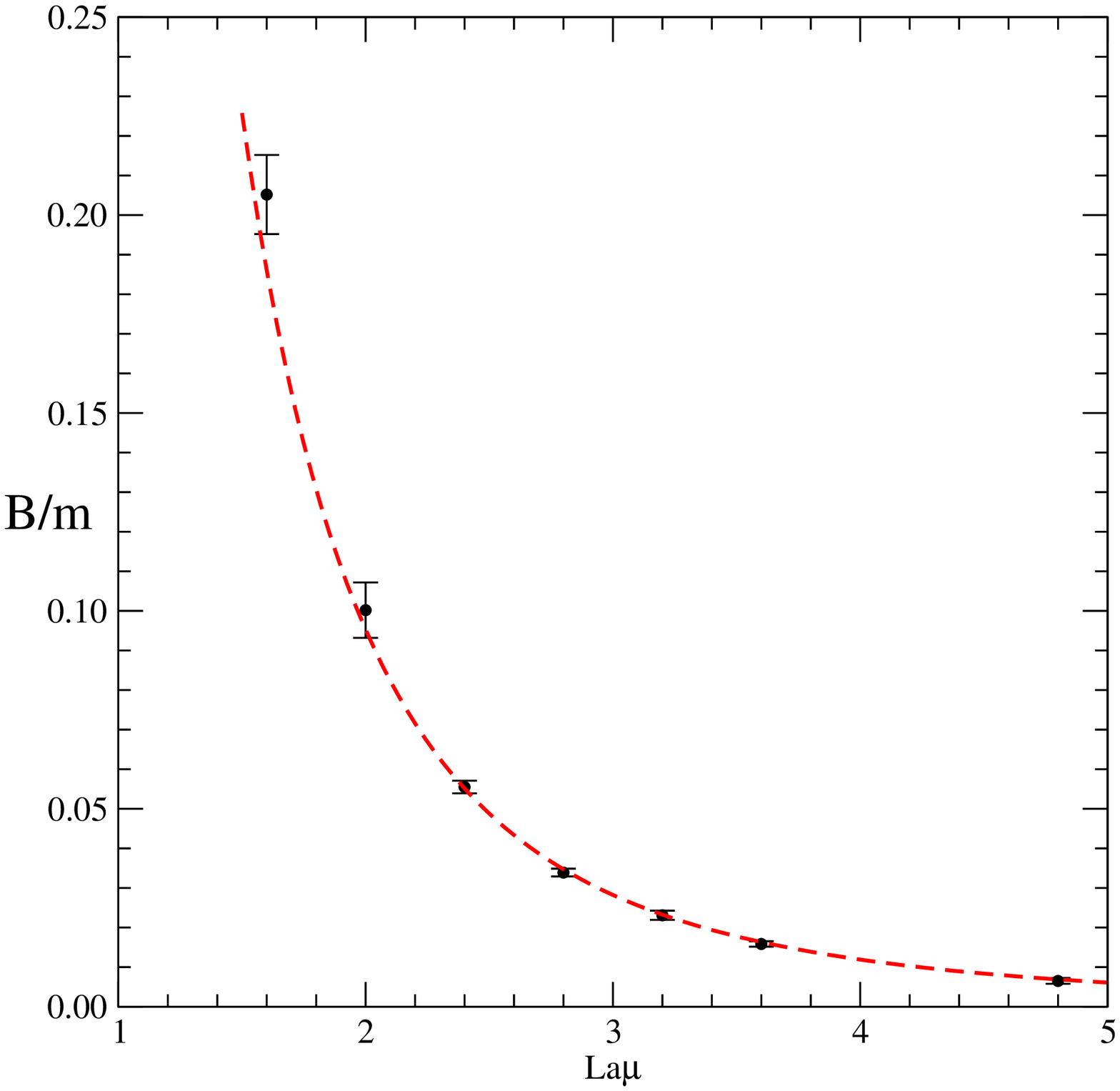}
\caption{Binding energy versus lattice size $L$, for $g_L=0.3$, $\kappa=0.118$, and $a\mu=0.1$ averaged over 4000 samples for $L=16,\cdots,32$, 2000 for $L=36$,
and 800 for $L=48$. Dotted line corresponds to a $1/L^3$ fit. }\label{fig:B_vs_L}
\end{center}
\end{minipage}
\hspace{0.5cm}
\begin{minipage}[h!]{7.cm}
\begin{center}
\includegraphics[width=6.9cm]{a0mu_vs_G_new.eps}
\caption{Scattering length vs $G$ for a
lattice volume $La\mu=2.4$. Solid line indicates the continuum non-relativistic
result and the dotted one the Born approximation (\ref{Born}). Blue circles represent the NR results. }\label{fig:a0_naive}
\end{center}
\end{minipage}
\end{center}
\end{figure}

It was found that such ``ill-conditioned configurations'' appear in the Yukawa model
for almost any $\kappa$ when $g_L \gtrsim 0.6$. In this case the inversion of the Dirac operator
becomes in practice impossible.
For illustrative purposes, we have plotted in Fig. \ref{fig:cn}  the condition number
of $D$ as a function of the lattice coupling constant $g_L$ for an ensemble of $L=8$
configurations at fixed value of $\kappa$.  As one can see,  the condition number of a given
configuration diverges on a discrete set of $g_L$ values for $g_L \gtrsim 0.6$ indicating
the practical impossibility to compute the nucleon propagator.
The precise $g_L$ values where this divergence occurs depend on the particular
configuration, on the values of $\kappa$ and $a\mu$  and  on the lattice size.
It turns out however that the situation described in Fig.  \ref{fig:cn} is generic
for the quenched Yukawa model.

It is worth noticing that in the full  QFT formulation every configuration is weighted
by the determinant of the Dirac operator $D$ and therefore the configurations yielding an
ill-conditioned linear system (\ref{DS}), i.e with ${\rm det}(D)\approx0$, do not contribute to the
functional integral.  In the quenched  approximation, however, this is no longer true
and ``ill-conditioned configurations'' can be sampled.

Therefore, the numerical simulations in the quenched Yukawa model
are limited to values  of the lattice coupling constant $g_L\lesssim 0.6$.
Using a typical value of $\kappa=0.1$, this
corresponds  to $g={g_L\over2\kappa}\lesssim 3$, that is $\alpha={g^2\over4\pi}\lesssim 0.7$
which is of the same order than the $\alpha_{\rm QCD}$  in the non-perturbative region.

Although assuming that this  problem could be associated
to the quenched approximation it is physically surprising  that no any NN bound state could be generated if the N\={N} pair creation is not taken into account.

In absence of bound states  in the quenched approximation, we still can  access
to  the NN low energy scattering parameters.
The scattering observables cannot be obtained in Euclidean
time in the infinite volume limit \cite{MT_NPB_245_90} but can be extracted from  the volume dependent binding energy  measured on
finite lattices, like for instance the one plotted in Fig. \ref{fig:B_vs_L}.
The underlying formalism was developed by Luscher in \cite{ML_CMP104_86} who gave a $1/L$ expansion of the
binding energy. In its leading order  it reads:
\begin{equation}
\frac{B}{m} = - \frac{4\pi a_0\mu}{\left(\frac{m}{\mu}\right)^2 (La\mu)^3}  \label{a0naive}
\end{equation}
Using the binding energy values of Fig. \ref{fig:B_vs_L} and equation
(\ref{a0naive}), the NN scattering lengths $a_0$ can be computed.
The result --  corresponding to  $g_L=0.3$,  $\kappa=0.118$, and $a\mu=0.1$  -- is $a_0\mu\approx - 0.13$
and the dimensionless coupling constant of the non-relativistic model is $G=0.193$.
The corresponding non-relativistic scattering length value, given by Fig. \ref{lambda_G},  is $A_0=a_0\mu=-0.214$.

This study has been performed for several values of $g_L$. The dependence
of $a_0$ on the coupling constant $G$ is plotted in Fig. \ref{fig:a0_naive},
for a lattice size of $La\mu=2.4$ ($L=24$,  $a\mu=0.1$).
One can see that the lattice results notably departs
from the non-relativistic ones (solid line) and are above the Born approximation (dashed line).

The values of the accessible coupling constants
extend beyond  the Born regime but are still far from the pole
behavior corresponding to the appearance of the first bound state displayed in Fig. \ref{lambda_G}.
The difference between the lattice and NR results may indicate strong repulsive corrections.
These kind of corrections were already manifested in the bound state problem
when solving the same Yukawa model both in Light Front  and  BS  ladder equations.

\section{Conclusion}

We have presented the results of the Yukawa model for the NN system using three different
dynamical frameworks.
In this model -- which constitutes the simplest renormalizable quantum field theory of a  fermion-meson interaction --
two fermions of identical mass $m$ interact by exchanging a massive scalar particle of mass $\mu$.
Results are restricted to the $J^{\pi}=0^+$ state.

\bigskip
In non-relativistic dynamics the model consists in solving the Schrodinger equation
with the so called -- static and local -- Yukawa  potential (\ref{VYuk_r}).
The results depend on three parameters ($m,\mu,g$) but some simple scaling properties make it dependent
on a single dimensionless parameter, the coupling constant  $G=(g^2/4\pi)(m/\mu)$.
The existence of a first bound state requires the coupling constant to be greater than a  critical value $G_0=1.680$.
For $G>G_0$, its binding energy increases  monotonously and can reach any arbitrary value, even greater than $2m$.

\bigskip
The way for implementing relativistic dynamics in the description of the same system is not unique.
We have considered two relevant relativistic equations: Light-Front and Bethe-Salpeter.
In both cases, for this particular coupling and state, the corresponding equations can be integrated without
introducing any regularization form factor.
The scaling properties of the non-relativistic model model are now lost
and both equations exhibit  a critical coupling constant $g_c$, for  which  $M^2$ tends to $-\infty$.
For slightly smaller value of $g$ the mass  $M^2$ crosses zero: the system  "collapses".
The precise numerical value of $g_c$ depends however on the particular dynamical framework: $g_c=6.84$ for Light-Front equation
and $g_c=2\pi$ for {BS} one.
This value is however large enough to generate several bound states.
The corresponding relativistic binding energies $B(g^2)$ are close to each other but
they depart sizeably from the non-relativistic ones even for very loosely bound states and are strongly repulsive.

\bigskip
Relativistic equations are the first step  towards a  full Quantum Field Theory solution.
They suffer from two main drawbacks.
On one hand, most of the one-boson exchange kernels require to be regularized in order to obtain an integrable equation.
This is usually done by introducing a vertex form factor cutting the high momentum
components above some arbitrary value $\Lambda$,  but thus diluting all the benefit of  an approach starting from
the first principles, like underlying Lagrangian.
The Yukawa model (in $J=0^+$ state)  is rather an exception than the generic case of one-boson exchange models.
On the other hand, the ladder kernel accounts only for a small part of the interaction, specially  when large values of the coupling constant are involved.
We have presented the first attempt to incorporate the full dynamical content of the Yukawa model
by using standard lattice techniques developed in the context of QCD.

The only approximation in the calculations was to neglect the \={N}N loops (quenched approximation) as it is the case
in all the nuclear models.
The lattice results indicate the existence of a maximal coupling constant $g_L$ which is well below  the required value for
the existence of the first bound state in the heory.
For smaller coupling constant the scattering length has been computed using Luscher's prescription.
Results are in agreement with the repulsive effect found by solving the relativistic equations.

\bigskip
We have shown how the same interaction model  between fermions can give rise to very different physical  pictures
depending on the particular dynamical framework in which it is considered.
The differences between a relativistic and a non-relativistic  approach are not only quantitative but lead to new qualitative behavior.
In the particular case of the Yukawa model, which was the first of all the nucleon-nucleon models, the full quantum field solution
remains still unknown.


\end{document}